\documentclass[11pt]{article}
\usepackage{moriond,epsfig}

\def\Journal#1#2#3#4{{#1} {\bf #2}, #3 (#4)}

\def\be{\begin{equation}}
\def\ee{\end{equation}}
\def\bea{\begin{eqnarray}}
\def\eea{\end{eqnarray}}

\begin{document}
\vspace*{4cm}
\title{High Energy Radiation from Neutron Star Binaries}

\author{T. Di Salvo}

\address{Astronomical Institute "Anton Pannekoek," University of 
              Amsterdam, Kruislaan 403, NL 1098 SJ Amsterdam, 
              the Netherlands}

\author{L. Stella}

\address{INAF - Osservatorio Astronomico di Roma, Via Frascati 33, 
             00040 Monteporzio Catone (Roma), Italy}

\maketitle\abstracts{
This paper surveys our current knowledge of the hard X-ray emission properties 
of old accreting neutron stars in low mass X-ray binaries. Hard X-ray 
components extending up to energies of a few hundred keV have been clearly 
detected in sources of both the Atoll and Z classes. The presence and 
characteristics of these hard components are discussed in relation 
to source properties and state. An overall anticorrelation between the 
fraction luminosity in hard X-rays and mass accretion rate is apparent over 
different sources spanning a large range of luminosities as well as individual 
source undergoing state changes.
Evidence for a second, yet unknown, parameter controlling the hard X-ray 
emission is emerging.
We draw a parallel with the spectral properties of  X-ray binaries hosting a 
stellar mass accreting black hole, and conclude that, at a merely 
phenomenological level, there appears to be a close analogy between the 
spectral properties of black hole candidates in their high and 
intermediate states and Z-sources. We briefly mention models that have 
been proposed for the hard X-ray emission of neutron star low mass 
X-ray binaries and comment on perspectives in the INTEGRAL era.}

\section{Introduction}

A variable hard component dominating the spectrum of Sco X-1 above 
$\sim 40$~keV was detected as early as 1966 (Peterson \& Jacobsen 1966; 
see also Riegler et al. 1970; Agrawal et al. 1971; Haymes et al. 1971). 
In other occasions the hard tail in Sco X-1 was not found, perhaps owing
to pronounced variations (e.g., Miyamoto \& Matsuoka 1977, and references
therein; Soong \& Rothschild 1983; Jain et al. 1984; Ubertini et al.
1992). 
Evidence for a hard component was also found in Cyg~X-2 (Peterson 1973) and
GX~349+2 (Greenhill et al. 1979). These results received relatively little 
attention, probably because the nature of Sco~X-1-like and bright galactic 
bulge X-ray sources remained not understood or, at least, controversial 
until late seventies. 
On the contrary, that Cyg~X-1 hosts an accreting black hole candidate, BHC, had 
become clear as early as 1972 (Bolton 1972; Webster \& Murdin 1972).
The conspicuous hard X-ray emission of this source (see e.g.\ Tanaka \& 
Lewin 1995, and references therein) was therefore considered the prototypical 
"hard spectrum" of an accreting black hole and provided much of the 
observational basis for model development.
Optically thick, geometrically thin accretion disk models that 
were developed in those years proved inadequate to explain the hard 
power-law like spectrum of Cyg~X-1, that extended without a break up to 
$\sim 80-100$~keV in the hard state and was (likely) detected up to 
energies of $\sim 1$~MeV in soft/intermediate states (e.g.\ Liang 
\& Nolan 1984). 
Standard accretion disk models were thus modified to include the
presence of a hot inner disk region or corona, where unsaturated thermal 
Comptonisation of soft photons from the optically thick disk up to 
energies of many tens or hundreds keV takes place (Eardley, Lightman, \& 
Shapiro 1975; Eardley \& Lightman 1976; Galeev, Rosner \& Vaiana 1979). 

Renewed interest in the hard X-ray emission properties of neutron star low 
mass X-ray binaries was motivated by the SIGMA/GRANAT discovery of a 
spectral component extending up to energies of $\sim 100-200$~keV
in Terzan 2 (Barret et al. 1991), KS 1731-260 (Barret et al. 1992), SLX 
1735-269 (Goldwurm et al. 1996) and Terzan 1 (Borrel et al. 1996). 
Unlike Sco~X-1 (and similar sources), the sources above have relatively low 
luminosity ($\sim 10^{36}-10^{37}$ ergs/s) and emit type I X-ray bursts;  
moreover some of them are transients.

\section{Neutron star low mass X-ray binary basics}

Low mass X-ray binaries, LMXBs, contain neutron stars that have been spun up 
to periods of a few milliseconds by accretion torques (see e.g. van der klis 
2000; Strohmayer 2001). The fastest spin periods inferred so far are 
$\geq 1.5$~ms, i.e. longer than the mass shedding limit for 
virtually any equation of state. Direct evidence for the presence of a 
magnetosphere has so far been found in SAX J1808.4--3658, XTE J1751--305
and XTE J0929--314, the only three LMXBs displaying coherent pulsations in 
their quiescent emission (spin periods of 2.5, 2.3 and 5.4 ms, respectively,
Wijnands, \& van der Klis 1998; Markwardt et al. 2002; Galloway et al. 2002). 
The neutron star magnetic field is estimated to be $\sim 10^8-10^9$~G in 
these sources; other LMXBs which do not show coherent pulsations and have 
comparable (or higher) luminosities likely have comparable (or lower)
magnetic field strengths. 

The modern classification of LMXBs relies upon the branching displayed by 
individual sources in the X-ray color-color diagram assembled by using the 
sources' count rate over a "classical" X-ray energy range 
(typically 2-20 keV).  
This classification has proven very successful in relating the spectral and 
time variability properties (Hasinger \& van der Klis 1989; for reviews 
see van der Klis 1995, 2000; for recent work see Muno, Remillard, \& 
Chakrabarty 2002; Gierlinski \& Done 2002; Barret \& Olive 2002) depending 
on the pattern described by each source in the X-ray color-color diagram. 
It comprises a Z-class (source luminosities close to the Eddington 
luminosity, $L_{\rm Edd}$) and an Atoll-class (luminosities of 
$\sim 0.01-0.1\ L_{\rm Edd}$).
Most Atolls emit Type-I X-ray bursts, i.e. thermonuclear flashes in the layers 
of freshly accreted material onto the neutron star surface; only two 
Z-sources are (somewhat peculiar) X-ray bursters. 
Considerable evidence has been found that the mass accretion rate 
(but not necessarily the X-ray luminosity) 
of individual Z-sources increases from the top left to the 
bottom right of the Z-pattern ({\it e.g.} Hasinger et al. 1990), 
i.e. along the so called horizontal, normal and flaring branches 
(hereafter HB, NB and FB, respectively). Similarly in Atoll sources 
the accretion rate increases from the so-called island to the top of the 
upper-banana branch. 

Attempts at decomposing the X-ray spectra of Z sources in terms of 
two (or more) components have adopted different approaches and the 
origin of the spectral components is still debated ({\it e.g.} Mitsuda et al. 
1984; White et al. 1986; 1988; Psaltis, Lamb, \& Miller 1995). 
Over the classical X-ray range, two main models were discussed and applied.
In the western model the spectrum is decomposed in the sum of an 
unsaturated Comptonised spectrum, supposedly produced in an
inner disk corona, plus a blackbody originating from close to the neutron star 
surface or the boundary layer between the disk and the neutron star (White et 
al. 1986; White, Stella \& Parmar 1988); in the eastern model the spectrum 
consists of the sum of an optically thick multi-temperature disk model 
(locally emitting like a pure blackbody) plus a blackbody again originating 
from the neutron star or the boundary layers (Mitsuda et al. 1984; Mitsuda
et al. 1989).

Both decompositions proved adequate for fitting the $\sim 1-20$~keV 
spectra of Z-sources (see, however, Hasinger et al. 1990: analyzing
Ginga data of Cyg X-2, they found that the western model was systematically
below the high energy data in a few spectra in the FB; in those cases
the eastern model gave better fits). Typical temperatures 
for the blackbody component are in $\sim 1-2.5$~keV range in the western 
and eastern model. The fraction 
of $1-20$~keV luminosity in the blackbody component is $\leq 20$\% 
in most cases. This value is substantially lower than that expected 
from boundary layers emission ($\geq 50$\% unless the neutron star 
rotates very close to mass shedding limit). 

Concerning Atoll sources, the eastern model could not adequately fit the 
spectra of some of the lower luminosity sources, where the spectrum extended 
without any evidence for a break up to highest measured energies 
(White, Stella \& Parmar 1988).
In essence, this is due to the fact that given the local blackbody emissivity 
of the multitemperature disk model, the spectrum consist of a power law 
with photon index $\Gamma = -2/3$ up to an exponential cutoff 
corresponding to the temperature of the innermost (and hottest) disk 
region. Therefore the eastern model was modified in order to include the 
effects of the Comptonisation of the $\sim 2$ keV blackbody coming from the 
neutron star (Mitsuda et al. 1989).
%%An additional (power law-like) component was therefore added within 
%%the eastern model in order to fit the spectra of the hearder Atoll sources.
In the unsaturated Comptonised component of the western model, the power law 
slope and cutoff energy are essentially determined by the Thomson depth and 
electron temperature of Comptonising region. These parameters can vary over 
a wide enough range to fit the $\sim 1-20$~keV spectra of the harder Atoll 
sources as well. Moreover, we note that in some Atoll sources the additional 
blackbody component is not required within either of the spectral 
decompositions. 

%A tracking of the spectral variations along the Z pattern was attempted 
%for selected sources ; the 
%results have not proven conclusive so far, since in different decompositions 
%the spectral variations are complex and can be ascribed to different 
%components.
Attempts at ascribing spectral variations along the Z or Atoll pattern 
to variations of a single spectral component gave inconclusive results
(see {\it e.g.} Hasinger et al. 1990; Hoshi \& Mitsuda 
1991; Schulz \& Wijers 1993; Asai et al. 1994, and references therein).
Similarly, constructing a parallel between the spectral components of 
Z and Atoll sources proved difficult, especially for the eastern model
(Mitsuda et al. 1989).
By contrast, the timing properties of LMXBs, as inferred from different 
power spectrum components such as QPOs, noise components etc., show
a great deal of continuity along the different branches of each source and 
also a remarkable similarity across Z and Atoll sources (Wijnands \& 
van der Klis 1999; Psaltis, Belloni, \& van der Klis 1999; van der Klis 2000; 
Belloni, Psaltis \& van der Klis 2002).

\section{Hard X-ray spectral components in Atoll sources}

\begin{figure}
\hspace*{-1.3cm} \psfig{figure=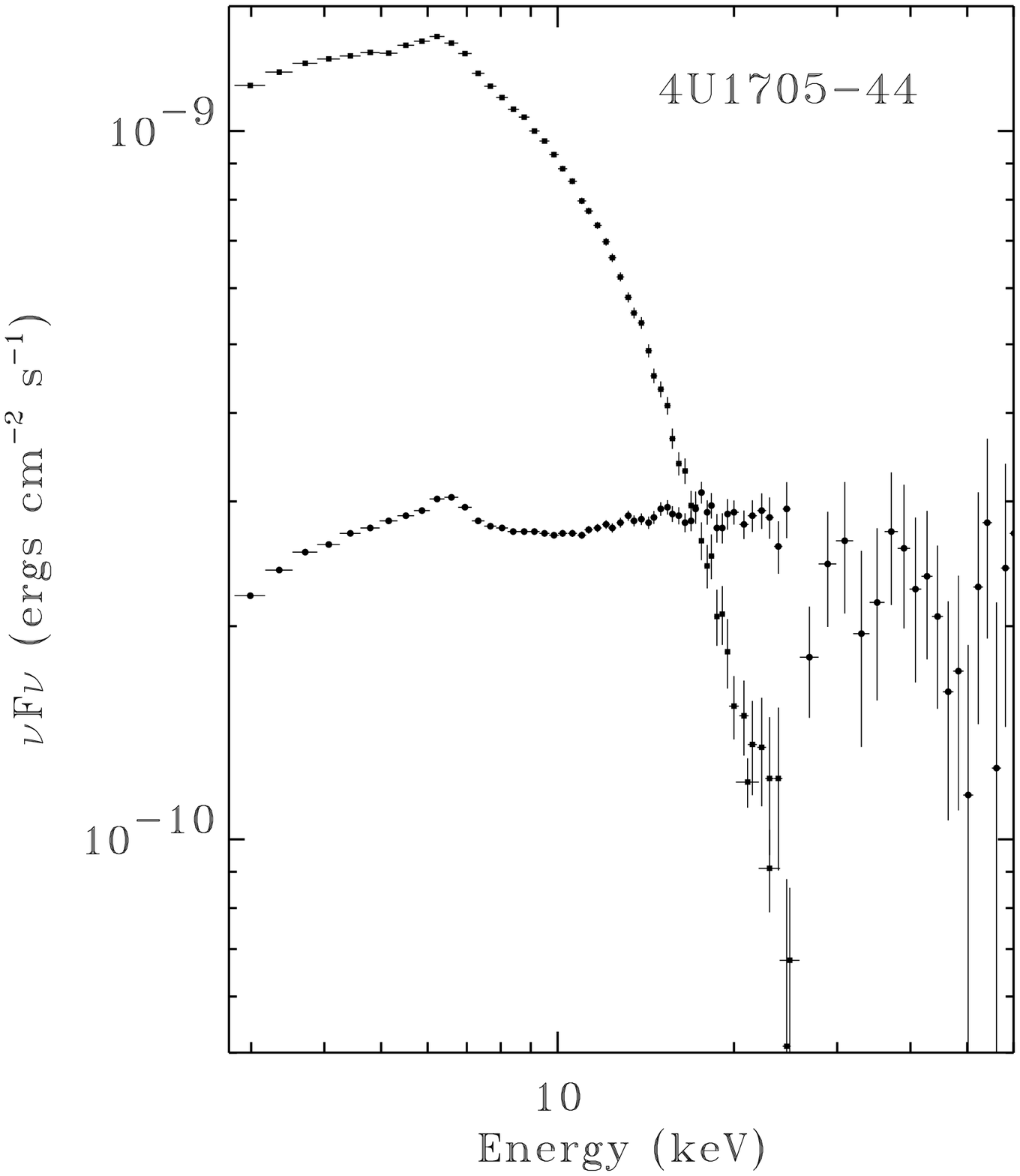,height=3.4in}
\hspace*{-1.3cm} \psfig{figure=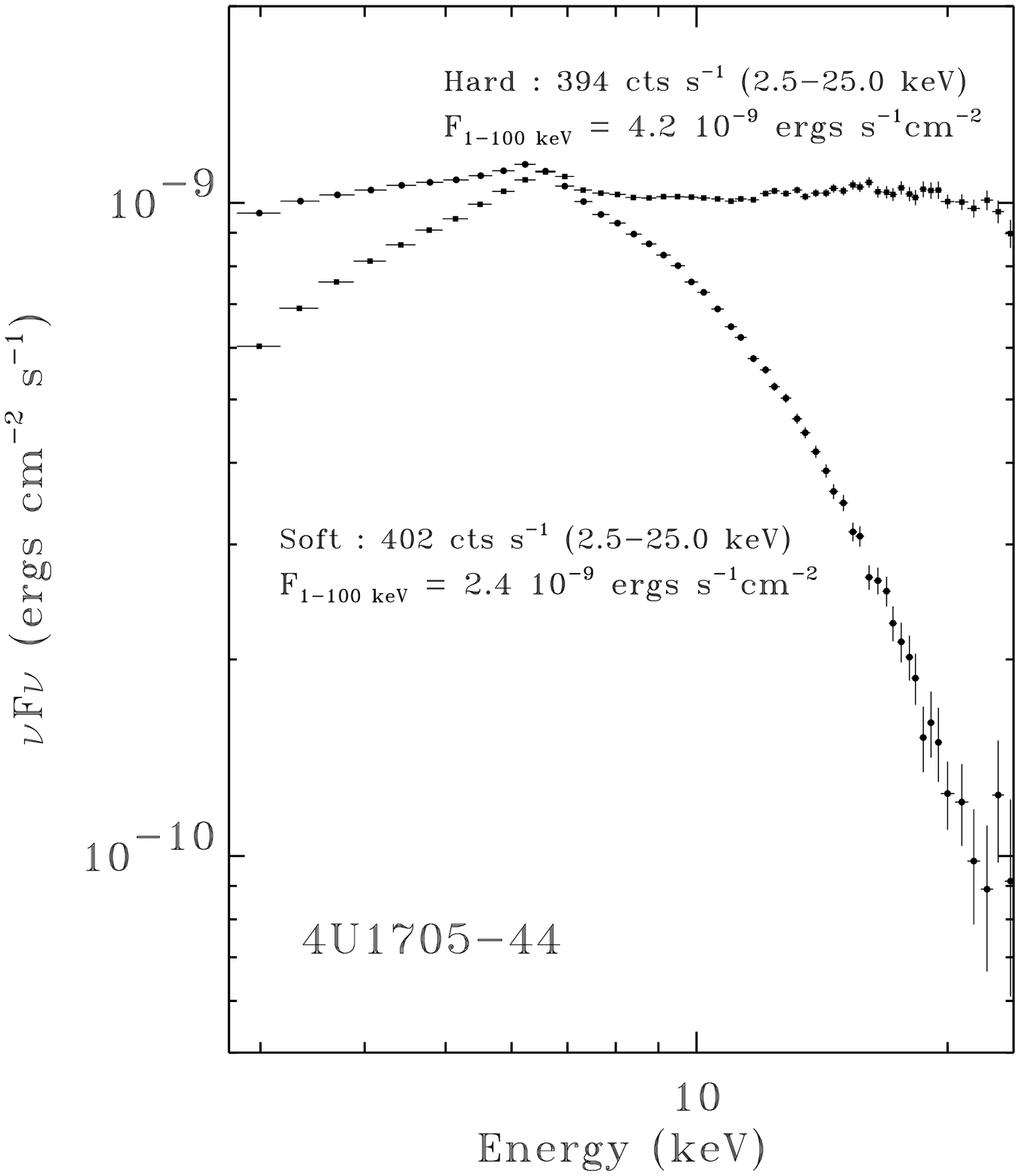,height=3.4in}
\caption{{\it Left)} Soft and hard spectral states from the LMXB 4U 1705-44 
  as observed by RXTE (PCA and HEXTE spectra are combined). 
  {\it Right)} PCA spectra taken by RXTE from 4U 1705-44 during a spectral 
  transition that occurred in February 1999 (Barret \& Olive 2002, see also 
  Barret 2001).  The broad band luminosity in the hard 
  spectrum is about twice the one associated with the soft spectrum, because 
  of the presence of a strong hard X-ray component. This illustrates that the 
  X-ray count rate alone is not a good indicator of the spectral state.
\label{fig1}}
\end{figure}

Hard X-ray components extending up to energies of several hundred 
keV have been revealed in about 20 neutron star LMXBs of the Atoll class
(some recent observations of hard X-ray spectra from low-luminosity LMXBs 
are reported in Table~\ref{tab1}). 
\begin{table}[h!]
\small
\begin{center}
\caption{Some recent observations of hard X-ray spectra from low-luminosity LMXBs.}
\vskip 0.5cm
\label{tab1}
\begin{tabular}{l|c|c|c|c|c} 
\hline \hline
Source 		  &        Luminosity        & Photon index & $kT_e$ & $E_{\rm max}$  & Ref. \\
       		  & ($10^{37}$ ergs/cm$^2$/s)&              & (keV)  & (keV)          &      \\            
\hline
4U 0614+09	  & 0.5--2.9 (0.1--200 keV)  & 2.3--2.4	    & $> 150$& 200	    
& \cite{pir99}    \\
4U 1608--52 	  & 0.3--1 (2--60 keV)       & 1.7--2.2	    & 30--60 & 200	    
& \cite{yos}, \cite{zha96}  \\
4U 1705--44	  & $\sim 1.5$ (1--20 keV)   & 1.5	    &$\sim25$&  20	    
& \cite{bar94}$^a$\\
Ter 2 (X 1724-308)& $\sim 2$ (0.1--100 keV)  & 1.6--1.9     & 30--90 & 200	    
& \cite{bar94}, \cite{bar91}, \cite{gua98}, \cite{bar00} \\ 
MXB 1728--34 (GX 354-0)&$\sim0.7$ (35--200 keV)&$3.0\pm0.2$ & --     & 200	    
& \cite{bar94}    \\ 
KS 1731--260	  & $\sim 1$ (35--150 keV)   &$2.9 \pm 0.8$ & --     & 150	    
& \cite{bar92}, \cite{bar94}  \\
Ter 1 (X 1732-304)&$(4.0 \pm 0.8) \times 10^{-2}$ (40--75 keV)
					     & $3.2\pm 0.7$ & 20--60 & 170
& \cite{bor}	  \\					     
SLX 1735--269	  & $\sim 1.3$ (1--20 keV)   & 2--3         & 30--50 & 200	    
& \cite{bar00}, \cite{gold}    \\
SAX J1747.0--2853 & $\sim 0.26$ (2--10 keV)  & 1.6--2	    & 30--70 & 200	    
& \cite{nat00b}   \\
SAX J1748.9--2021 & $\sim 1.2$ (0.1--200 keV)& $1.44\pm0.5$ & 20--50 & 100	    
& \cite{int99a}   \\
SAX J1808.4--3658 & $\sim 0.38$ (2--200 keV) & $1.82\pm0.04$&$180^{+120}_{-60}$& 200
& \cite{gie02b}   \\
SAX J1810.8--2609 & $\sim 0.12$ (1--200 keV) & $1.96\pm0.04$& --     & 200	    
& \cite{nat00a}   \\
4U 1820--30	  & 2.3--2.6 (2--50 keV)     & $2.05\pm0.05$&$\sim 20$& 50	    
& \cite{bar94}$^a$, \cite{blo00b} \\
GS 1826--238 	  & $\sim 0.88$ (1--20 keV)  & $\sim 1.7$   &$\sim 90$& 150	    
& \cite{bar00}, \cite{int99b} \\
X 1850--087	  & $\sim 0.13$ (1--37 keV)  & $\sim 2.3$   & --     &  37	    
& \cite{bar94}$^a$\\
4U 1915--05	  & $\sim 0.6$ (2--50 keV)   & 1.8--1.95    & $>100$ & 200	    
& \cite{blo00}$^a$\\
Aql X-1		  & $\sim 0.2$ (1--100 keV)  & 2.2--2.6	    & --     & 150	    
& \cite{har96}    \\
\hline
\end{tabular}
\end{center}
Luminosity is the source luminosity calculated
in the energy range specified in brackets; $kT_e$ is the 
inferred electron temperature in the hot Comptonizing corona;
$E_{\rm max}$ is the maximum energy at which the source was 
observed. \\
$^a$ and references therein. 
\end{table} 
In these systems the power law-like component, with typical slopes 
of $\Gamma \sim 1.5 -2.5$, is followed by an exponential cutoff,
the energy of which is often in between $\sim 20$ and many tens of keV.
This component is interpreted in terms of unsaturated thermal Comptonisation. 
There are instances in which no evidence for a cutoff 
is found up to $\sim 100-200$~keV. This is the so called "hard state"
of Atoll sources. There are sources that appear to spend most of the time 
in this state (e.g.\ 4U 0614+091, Ford et al. 1996, Piraino et al. 1999, and
references therein). In others a gradual transition from the soft to the 
hard state has been observed in response to a decrease of the source 
X-ray luminosity and/or the source drifting from the banana branch 
to the island state (see Fig.~\ref{fig1}, left panel).
This transition is often modelled in terms of a gradual decrease of the 
electron temperature of the Comptonising region. 
%%We note that the within the Western model, the hard state spectra of 
%%Atoll source does not usually require the addition of another spectral 
%%component; this is instead the case in the Eastern model (see also section 2).
%%

A reflection component in the characteristic shape of a broad bump centered 
around energies of a few tens of keV, sometimes together with a Fe K-shell 
line and edge, was detected in the hard state of some Atoll sources 
(e.g.\ Yoshida et al. 1993; Piraino et al. 1999; 
Barret et al. 2000). This component is believed to originate from reflection
of the optically thick disk regions. 
Usually the reflection amplitudes ({\it i.e.} 
The solid angle $\Omega/2\pi$ subtended by the reflector as seen from the 
hot inner corona  are lower than 0.3. A correlation has been claimed between 
the photon index of the primary spectrum and the reflection amplitude of the
reprocessed component (Zdziarski et al. 1999; Piraino et al. 1999; 
Barret et al. 2000), similar to that reported for 
BHCs and Seyfert galaxies.

As first noted by van Paradijs \& van der Klis (1994), there is a  
clear trend for the spectral hardness of these sources (and accreting 
X-ray sources in general) over the 13-25 and 40-80~keV energy ranges to be 
higher for lower X-ray luminosity.
Therefore mass accretion rate appears to be the main parameter 
driving the spectral hardness of Atoll sources, both as a group and as 
individual sources. 
Yet there is evidence that at least on occasions an additional parameter 
controls the soft/hard spectral transitions. This is especially apparent 
from a recent study of 4U 1705--44, in which the source underwent a soft to 
hard state transition while the 0.1-200~keV bolometric luminosity of the 
source decreased by a factor of $\sim 3$ from the soft to the hard state 
and increased by only a factor of $\sim 1.2$ in the opposite transition from 
the hard to the soft state (Fig.~\ref{fig1}, right panel; 
Barret \& Olive 2002).
On another occasion the same source displayed hard and soft states 
which were found to differ by a much larger factor 
(up to one order of magnitude) 
in their luminosity (Fig.~\ref{fig1}, left panel). 
It has been suggested that the second parameter regulating the spectral 
state transitions might be the truncation radius of the optically thick 
disc. However, what determines the radius at which the disc is truncated 
is not clear yet: this could be the mass accretion rate through the disk 
normalized by its own long-term average (as proposed by van der Klis 2001 
to explain the ``parallel tracks'' observed in the kHz QPO frequencies vs. 
X-ray flux diagram), but also magnetic fields or the formation of jets 
could play a role.

\section{Hard X-ray components in Z-sources}

Recent broad band studies have shown that many Z-sources  
display variable hard, power-law shaped components, dominating 
their spectra above $\sim 30$~keV. As mentioned in the introduction, hard 
tails in Z-source spectra were occasionally detected 
in the past. The first detection was in the spectrum of Sco X--1; beside the 
main X--ray component (equivalent bremsstrahlung temperature of $\sim 4$~keV) 
Peterson \& Jacobson (1966) found a hard component dominating the spectrum 
above 40~keV.  
The latter component was observed to vary by as much as a factor of 3. 
More recently the presence of a variable hard tail 
in Sco X--1 was confirmed by OSSE and RXTE observations
(Strickman \& Barret 2000; D'Amico et al. 2001).
%; surprisingly 
%no clear connection was found between the source 
%position along the Z-pattern and the intensity of the hard component.
%A hard tail was also 
%detected in 

Much progress in the study of the spectrum of Z-source and other 
high luminosity LMXBs 
has been recently achieved through BeppoSAX observations:
a hard X-ray tail was found in the Z-sources GX 17+2 (Di Salvo et al. 2000), 
GX~349+2 (Di Salvo et al. 2001) and Cyg~X-2 (Frontera et al. 
1998; Di Salvo et al. 2002), as well as the peculiar 
bright LMXB Cir X--1 (Iaria et al. 2001, 2002) and during type II 
bursts from the Rapid Burster (Masetti et al. 2000). 
The fact that a similar hard component 
has been observed in several Z sources indicates that this is 
probably a common feature of these sources.

\begin{figure}[h!]
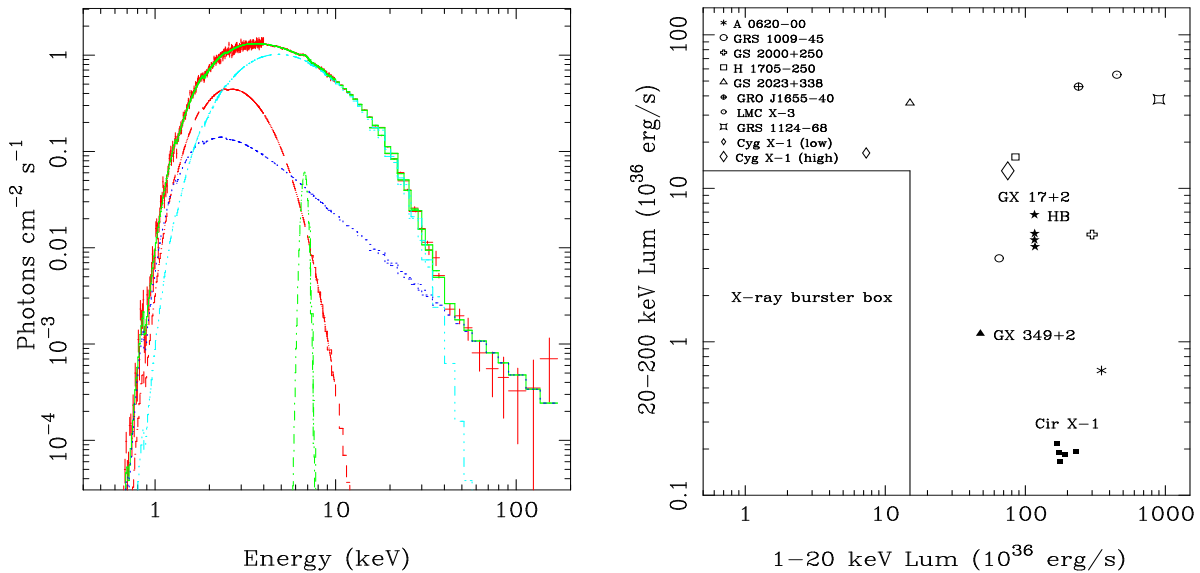

$$\psfig
{figure=fig3.ps,height=7.5cm,width=7.5cm}\qquad
\psfig{figure=fig4.ps,height=7.5cm,width=7.5cm}$$
\caption{{\it Left)} BeppoSAX photon spectrum of GX~17+2 
in the HB. The solid line represents the best fit model;
individual model components are also 
shown. The hard power law, dominating the spectrum
above 30 keV, is plotted as a dotted line. 
{\it Right)} 20-200 keV versus 1-20 keV luminosities of black hole binaries 
(open symbols, from Barret et al. 2000) and neutron star type-Z binaries 
(filled symbols, from Di Salvo et al. 2001).  The so-called {\it X-ray burster 
box} is plotted as a solid line. Its boundaries are defined as in Barret 
et al. (2000). 
% pensi che si possano aggiungere anche i punti delle Atolls  
% in ogni caso aggiungerei anche i punti di X1118 (BHC) che cadono 
% dentro la burster box ed aggiungerei il pezzo che ho messo nel 
% testo come footnote  
\label{fig2} \label{fig2b} }
\end{figure}

This hard component can be fitted by a power law, with photon index in the 
range 1.9--3.3, contributing up to 10\% of the source bolometric luminosity.
The parameters of these components, as deduced from recent X-ray observations,
are reported in Table~\ref{tab2}. 
\begin{table}[h!]
\small
\begin{center}
\caption{Recent observations of hard X-ray components in bright LMXBs.}
\vskip 0.5cm
\label{tab2}
\begin{tabular}{l|c|c|c|c|c|c} 
\hline \hline
Source & Spectral state & Luminosity              & Hard Luminosity         & Photon index & $E_{\rm
max}$ & Ref. \\
       &                &($10^{38}$ ergs/cm$^2$/s)&($10^{36}$ ergs/cm$^2$/s)&              &    (keV)
&      \\            
\hline
GX 17+2 &     HB        & 1.2                     & 2.0                     & $2.7 \pm 0.3$& 200
& \cite{dis00}    \\
GX 349+2&   NB/FB       & 0.49                    & 0.5                     & $1.9 \pm 0.4$& 200
& \cite{dis01}    \\
Cyg X-2 &     HB        & $0.89-1.1$              & 0.8--0.9                & $1.8-2.1$    & 200
& \cite{dis02}    \\                       
Sco X-1$^b$&  HB        & --                      & 0.9--1.4                & $1.7-2.4$    & 200
& \cite{dam}    \\    
        &     NB        & --                      & 0.8                     & $1.6 \pm 0.3$& 200
& \cite{dam}    \\
        &     FB        & --                      & 0.5--1.1                & $-1-0.1$     & 200
& \cite{dam}    \\
GX 5-1$^a$&   NB        & $\sim 3$                & $\sim 10$               & 1.8 (fixed)  &  37
& \cite{asa}    \\
Cir X-1 &   NB/FB       & 1.4                     & 0.16                    & $3.3 \pm 0.8$& 200
& \cite{iar01}    \\
        &  Unknown      & 0.97                    & 0.27                    & $3.0 \pm 0.4$& 200
& \cite{iar02}    \\               
\hline
\end{tabular}
\end{center}
Spectral in which the hard 
component was significantly detected: HB, Horizontal Branch, NB, Normal Branch, FB,
Flaring Branch; Luminosity is the (absorbed) luminosity 
in the 0.1--200 keV energy range; Hard luminosity is the luminosity in the 
power-law component in the 20--200 keV energy range; $E_{\rm max}$ is the 
maximum energy at which a source was detected.\\
$^a$ In this case a contribution from a nearby contaminating source could not 
be excluded.\\
$^b$ The 2--20 keV luminosity of Sco X--1 was $(2.1-3.2) \times 10^{38}$ 
ergs/cm$^2$/s.
\end{table} 
The presence of the hard component in Z sources is in some cases related to 
the source state or its position in the CD.  
This was unambiguously shown for the first time by the
BeppoSAX (0.1--200~keV energy range) observation of GX~17+2 
(Di Salvo et al. 2000, see Fig.~\ref{fig2}), where the hard tail was observed 
to vary systematically with the position of the source in the CD.
In particular the hard component (a power-law with photon index of $\sim 2.7$) 
showed the strongest intensity in the HB of its CD (see the corresponding 
spectrum in Fig.~\ref{fig2}); a factor of $\sim 20$ decrease was observed when 
the source moved from the HB to the NB, i.e.\ from low to high inferred 
mass accretion rate.  
In a {\it Ginga} (1.5--37~keV energy range) observation 
of GX~5--1 a hard excess was detected. This could be  
fitted by a power law with photon 
index 1.8, the intensity of which decreased from the NB to the FB (again 
from low to high mass accretion rate, Asai et al. 1994).
The hard component detected in BeppoSAX data of GX~349+2 (Di Salvo et al. 
2001) is among the hardest detected in bright LMXBs. 
Its photon index was of $\sim 1.9$, with no evidence of a high 
energy cutoff (up to $\sim 100$~keV). 
Cir X--1, thought to be a peculiar Z source (Shirey et al. 1999), 
was observed by BeppoSAX in the orbital phase interval 0.11--0.16, i.e.
close to periastron. The state of the source was identified with the
NB/FB (Iaria et al. 2001). Also in this case a hard tail was detected in 
the non-flaring spectrum, with a fairly steep power law slope 
of $\sim 3.3$. Cir X--1 was also observed by BeppoSAX at orbital phases
0.61--0.63, i.e.\ close to apoastron. Again a hard tail, with photon
index $\sim 3$, was required to fit its spectrum above 20 keV (Iaria et al.
2002).

In most of these cases the hard component becomes weaker for higher 
accretion rates (note that the contribution of the 
hard component to the total X-ray flux is different 
in different sources).  Yet, in recent HEXTE observations of
Sco X--1, a hard power-law tail was detected
in 5 out of 16 observations, without any clear correlation with the 
position in the CD (D'Amico et al. 2001). The behavior of Sco X--1 again
suggests that there might be a second parameter, besides 
mass accretion rate, regulating the presence of
hard emission in these systems. Interestingly, Strickman \& Barret (2000) 
suggest that the hard X-ray emission present in Sco X--1 data from OSSE may 
be correlated with periods of radio flaring.

\section{Comparison with Black Hole Binaries}

The hard X-ray spectra of Atoll sources in their hard (island) state are 
reminiscent of those of BHCs in their low state, LS. 
In both cases the hard component is well represented by a power-law  
spectrum that extends to many tens of keV; at higher energy an 
exponential cutoff sets in. Yet there are some notable differences. 
Firstly, the power-law slopes of the hard state Atolls are somewhat steeper
than those of LS BHCs (Barret \& Vedrenne 1994).
More crucially, in the transition to the soft (banana) state, 
i.e. higher accretion rates, the cutoff energy of Atoll sources decreases 
markedly; this behaviour is not observed in BHCs when these move from the 
LS to the intermediate state, IS, or very high state, VHS (where a steep
power law appears at high energies without evidences of a high energy 
cutoff).

On the other hand, the spectra of Z sources are quite similar to the spectra 
of BHCs in their soft states. At soft X-ray energies the latter, 
in fact, are dominated by soft emission with characteristic temperatures of 
$\sim 1-2$~keV, both in their IS and HS
(this is to be contrasted with the characteristic temperatures of 
$3-4$~keV, plus an additional softer blackbody-like component in Z-sources).

As already mentioned, a hard spectral component is apparent in the IS/VHS 
of BHCs, which can be modelled with a power-law with photon 
index $\sim 2-3$, extending up to many hundred keV and showing no clear 
evidence for a high energy cutoff (see e.g. Grove et al. 1998).
This component is very variable 
and usually contributes a few per cent of the total luminosity. 
This clearly parallels the characteristics 
of most Z-sources, which display a variable hard X-ray component in their 
HB and NB. At a purely phenomenological level, this is a clear indication 
that the presence of such a high energy tail in an otherwise soft 
spectrum cannot be considered a signature of black hole accretion.  
For the higher mass accretion rates of Z-sources (their FB) and in BHCs 
in the HS the intensity of the hard X-ray component is much reduced, often 
below detectability. 
Unlike BHCs, however, Z-sources have not yet shown transitions to a 
``low state'' dominated by a power-law like spectral component 
extending up to high energies. 

We note that even though mass accretion rate variations appear to be the main  
cause of the spectral transitions of BHCs, there is considerable evidence that 
a second, yet unknown, parameter can give rise to these transitions. 
The existence of a second parameter was indeed proposed to explain the 
soft/hard spectral transitions observed in the BHCs XTE~J1550--564 
(Homan et al. 2001) and GS2000+25 (Tanaka 1989).
%%The cases of XTE~J1550--564 (Homan et al. 2001) and GS2000+25 (Tanaka 1989)
%%are especially relevant in this respect.

The similarity between BHCs and Z-sources is also apparent from the 
``luminosity diagram'', in which the 1--20 keV luminosity is plotted
versus the 20--200 keV luminosity.   Barret et al. (1996) observed that
all (low-luminosity) neutron star systems of the Atoll class in which a hard 
component had been detected lie in what they called the ``X-ray burster box'', 
while all black hole systems lie outside.  
By plotting the X-ray luminosities 
of the Z-sources GX~17+2 (Di Salvo et al. 2000), Cir~X--1 (Iaria et al. 2001), 
and GX~349+2 (Di Salvo et al. 2001a) in the same diagram it is apparent 
that these sources lie in the region where BHCs lie, outside the X-ray 
burster box (see Fig.~\ref{fig2}, right panel).  

It appears, however, that only black hole candidates have a 
20--200 keV luminosity $\ge 1.5 \times 10^{37}$ erg/s; this might be related 
to the higher Eddington luminosity that characterizes BHCs.  
These similarities might indicate that the hard tails in both black holes 
and neutron stars originate from the same mechanism. This would imply that 
this mechanism does not require the presence of an event horizon.

There might exist a relationship between the presence of the hard X-ray 
component and QPOs.  Di Matteo \& Psaltis (1999) found an interesting 
correlation between the QPO frequency and the slope of the power-law energy 
spectra in BHCs: the photon index increases with increasing
the QPO frequency (see also Kaaret et al. 1998 for a similar correlation in 
Atoll sources). If the QPO frequency is related to the size 
of the innermost (optically thick, geometrically thin) disk region, which 
changes (mainly) in response to accretion rate variations 
(as indeed envisaged in a 
number of QPO models; see van der Klis 2000 for a review), then this 
correlation might simply reflect variations of the Comptonization parameter 
$y$ in response to variations of the innermost disk radius. 
In the case of Z sources, relatively poor statistics in the hard X-ray 
components prevented 
a study of the hard tail slope as the source moves from the HB to 
the NB, i.e., the path along which the QPO frequency increases.
Yet the data are consistent with the hypothesis of a correlation between
hard X-ray emission and QPO frequencies.
This is suggested by the fact that: (1) the hard X-ray component of GX 17+2 
is most pronounced over the source state(s) in which both the kilohertz 
and HB QPOs 
are detected and reach the highest rms amplitudes and (2) both the 
contribution from the hard X-ray component to the total spectrum and the 
rms amplitude of the QPOs increase dramatically with energy (e.g.\ 
van der Klis 2000). 

Especially interesting is the possible relationship between hard X-ray 
and radio emission.
All the Z sources are detected as variable radio sources. In the
case of Sco X--1, the brightest radio source among neutron star LMXBs,
recent VLBI observations have shown that the radio source consists of a 
variable core and two radio lobes which form close to the 
core and move outward at relativistic speeds, $\sim 0.45\;c$
on average (Fomalont et al.\ 2001a, 2001b). 
Usually the highest radio fluxes are associated with the HB. The radio 
emission weakens in the NB, and is not detected any longer in the FB
(e.g.\ Hjellming \& Han 1995, and references therein).
%In other words, 
Therefore, the radio emission from these objects, which probably
arises in the jets (Fender \& Hendry 2000), 
is anticorrelated with the inferred mass accretion rate.  
This seems to be a fairly general behaviour, holding for 
different kinds of accreting collapsed objects, BHCs, Z-source, as well as  
Atoll sources (although only a few Atolls have been detected in
radio so far, e.g.\ Fender 2001). 

Since the hard X-ray emission component from these sources is, in general, 
more pronounced for lower mass accretion rates, it 
has been proposed that non-thermal, high energy electrons, 
responsible for the hard tails observed in Z sources, might be accelerated 
in the jets (Di Salvo et al. 2000, Iaria et al. 2001a; see also Fender 2001).
%This hypothesis seems in general agreement with the observed behavior of 
%the hard X-ray emission, but further observations are needed to clarify 
%the correlation between hard X-ray and radio emission in these sources.
% ho tolto la frase sopra perche' mi sembrava un po' ``circolare'' 

\section{Models}

In this section we briefly summarize the main features of some 
of the models proposed for the hard X-ray emission of neutron star 
LMXBs (for more details and references see Barret 2001).  
The analogy between the hard X-ray spectral components of neutron 
star LMXBs and BHCs suggests a similar 
emission mechanism and geometry in all these systems.
The hard spectrum in neutron star systems (at least in 
those cases in which a thermal cutoff has been observed) can be explained 
as thermal Comptonisation of soft photons in a hot
region (corona), perhaps placed between the neutron star and the 
accretion disk. This is similar to models proposed for BHCs in their LS.

In some cases, however, a high energy cutoff is not observed up to energies
of $\sim 100$ keV or higher, which would imply extremely high electron
temperatures. In the case of Z sources these high temperatures would be 
difficult to explain considering that the most of the source emission 
sources is very soft.
As in the case of black hole candidates in the IS, the hard tails 
observed in Z sources can be produced either in a hybrid thermal/non-thermal 
corona (i.e.\ a corona in which a fraction of the energy can be injected 
in form of electrons with a non-thermal velocity distribution, Poutanen \& 
Coppi 1998) or in a bulk motion of matter close to the neutron star (e.g.\ 
Titarchuk \& Zannias 1998).  
Fast radial converging motions are unlikely to be dominant in the innermost 
region of the accretion flow in such high-luminosity systems, because of the 
strong radiation pressure emitted from near the neutron star surface.
However, power-law tails, dominating the spectra at high energy, can
also be produced when the flows are mildly relativistic ($v/c \sim 0.1$)
or when the velocity field does not converge (Psaltis 2001).  
Therefore outflows can be the origin of
these components, with flatter power laws corresponding to higher optical
depths in the scattering medium and/or higher bulk electrons velocities,
in a way that is similar to thermal Comptonization. 

An alternative possibility is that the hard X-ray component originates from 
the Comptonisation of seed photons by high-velocity electrons of a jet
(e.g.\ Di Salvo et al. 2000), or via optically thin synchrotron emission 
directly by the jet (Markoff, Falcke, \& Fender 2001).

\section{Perspectives in the INTEGRAL era}

Despite a considerable progress in recent years, the study of the 
hard X-ray emission properties from neutron star LMXB is still in its infancy.
Luminosity plots of the kind shown in Fig.~\ref{fig2} (right panel) is about 
as much as current observations can afford for a few tens of sources of this 
class. Hard X-ray color-color diagrams are still to come; the analogy with 
the color-color diagrams assembled from data in the classical X-ray 
band suggests that new interesting information will be derived from them. 
This will soon be made possible by INTEGRAL observations of a number 
of neutron star LMXBs. 

INTEGRAL holds also the potential to address key issues such as: 
(a) the origin (thermal or non thermal) of the hard
components in Z sources, which can be deduced, e.g., by the observation 
(or absence) of an exponential cutoff in the power-law hard tail;
(b) the correlation of the hard X-ray component with a 
variety of other source properties, such as source X-ray state, 
radio activity, X-ray line emission, QPOs frequencies etc. 
Coordinated observing programs exploiting the characteristics 
of INTEGRAL, as well as those of facilities such as Chandra, XMM and 
ground based optical and radio observatories will be especially relevant in 
this respect. 

%The broad 
%high energy coverage and good sensitivity of the $\gamma$-ray instruments 
%of the INTEGRAL mission, together with the X-ray monitor in the standard 
%X-ray range 3--35~keV, can represent an important step forward 
%in the understanding  of  
%the origin of high energy components in accreting X-ray binaries.

\section*{Acknowledgments}
We are grateful to our collaborators on these topics, L. Burderi, S. Campana,
R. Farinelli, F. Frontera, J. Homan, R. Iaria, G.L. Israel, E. Kuulkers, 
N. Masetti, A.N. Parmar, N.R. Robba, M. van der Klis, and to the Moriond 
meeting organizers. 
This study was partially supported through ASI grants and by the Netherlands 
Organization for Scientific Research (NWO).

\end{document}